\documentclass[12pt]{iopart}
\usepackage{iopams}
\usepackage{epsfig}
\begin{document}
\newcommand{\A}{{\rm A}}
\newcommand{\rpa}{R_{_{p\A}}}
\newcommand{\xt}{x_{_\perp}}
\newcommand{\ptrans}{p_{_\perp}}
\newcommand{\rag}{R^{^\A}_{_G}}
\newcommand{\rafd}{R^{^\A}_{_{F_2}}}
\newcommand{\dd}{{\rm d}}
\newcommand{\X}{{\rm X}\,}
\newcommand{\sqrtsnn}{\sqrt{s_{_{\mathrm{NN}}}}}
\newcommand{\rapprox}{R^{^{\rm approx}}}
\newcommand{\rppb}{R_{_{p {\rm Pb}}}}
\newcommand{\rpbpb}{R_{_{{\rm Pb}{\rm Pb}}}}

\begin{flushright}
  {\sffamily CERN-PH-TH/2007-119}; {\sffamily LAPTH-1197/07}
\end{flushright}

\title{Prompt photons in heavy ion collisions at the LHC: A ``multi-purpose'' observable}

\author{Fran\c{c}ois Arleo\footnote{On leave from Laboratoire d'Annecy-le-Vieux de Physique Th\'eorique (LAPTH), UMR 5108 du CNRS associ\'ee \`a l'Universit\'e de Savoie, B.P. 110, 74941 Annecy-le-Vieux Cedex, France}}

\address{CERN, PH department, TH division\\
1211 Geneva 23, Switzerland}

\begin{abstract}
I emphasize in this contribution how prompt photons can be used to probe nuclear parton densities as well as medium-modified fragmentation functions in heavy ion collisions. Various predictions in $p$--A and A--A collisions at LHC energies are given.
\end{abstract}

\begin{figure}[h]
  \hspace{-1.2cm}
  \begin{minipage}[t]{9.2cm}
    \begin{center}
      \includegraphics[height=6.9cm]{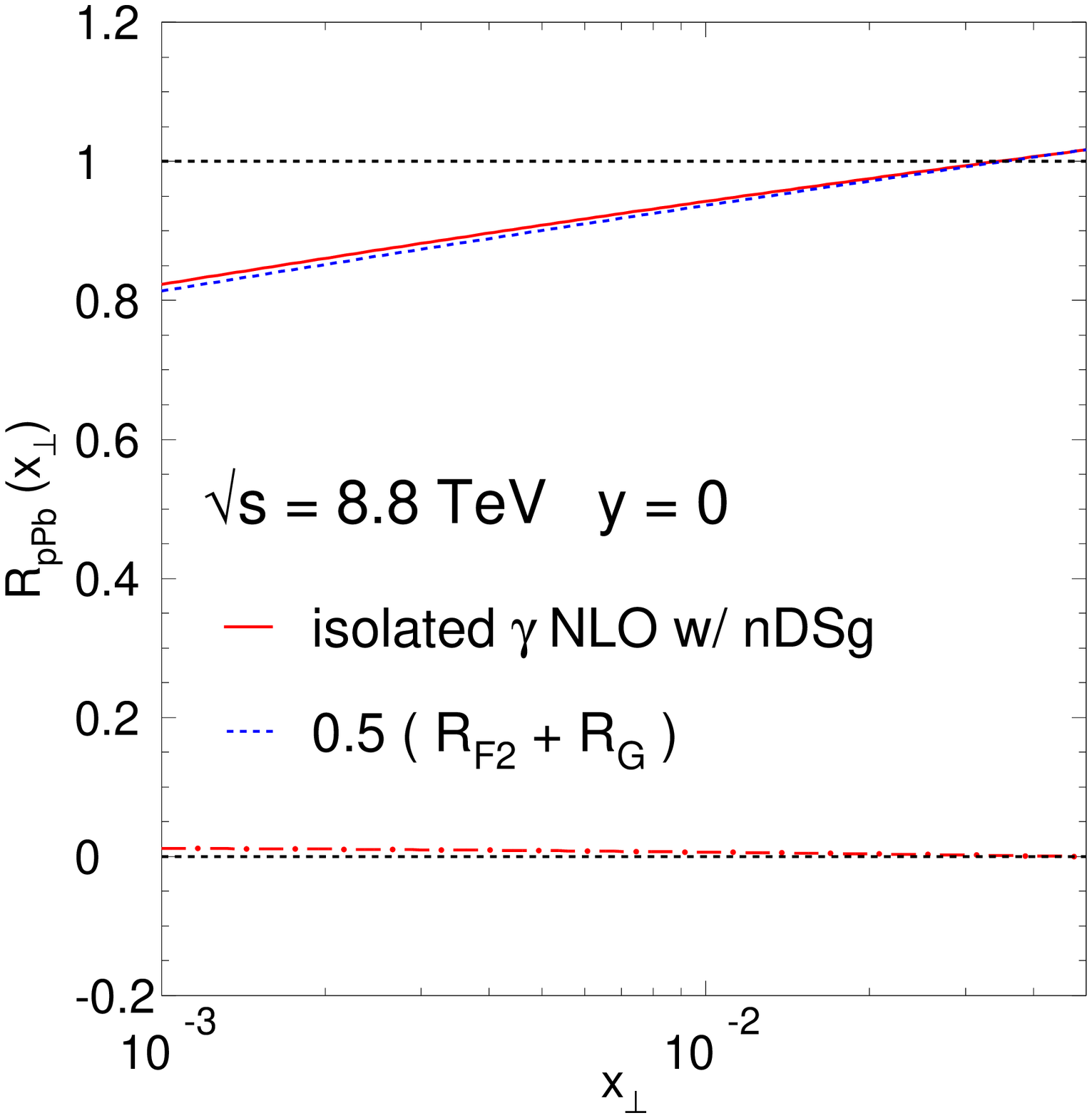}
    \end{center}
  \vspace{-0.5cm}
    \caption{$\rppb$ of $y=0$ isolated photons in $p$--Pb collisions at $\sqrtsnn=8.8$~TeV.}
    \label{fig:isoy0lhc}
  \end{minipage}
  \hspace{-1.8cm}
  \begin{minipage}[t]{9.2cm}
    \begin{center}
      \includegraphics[height=6.9cm]{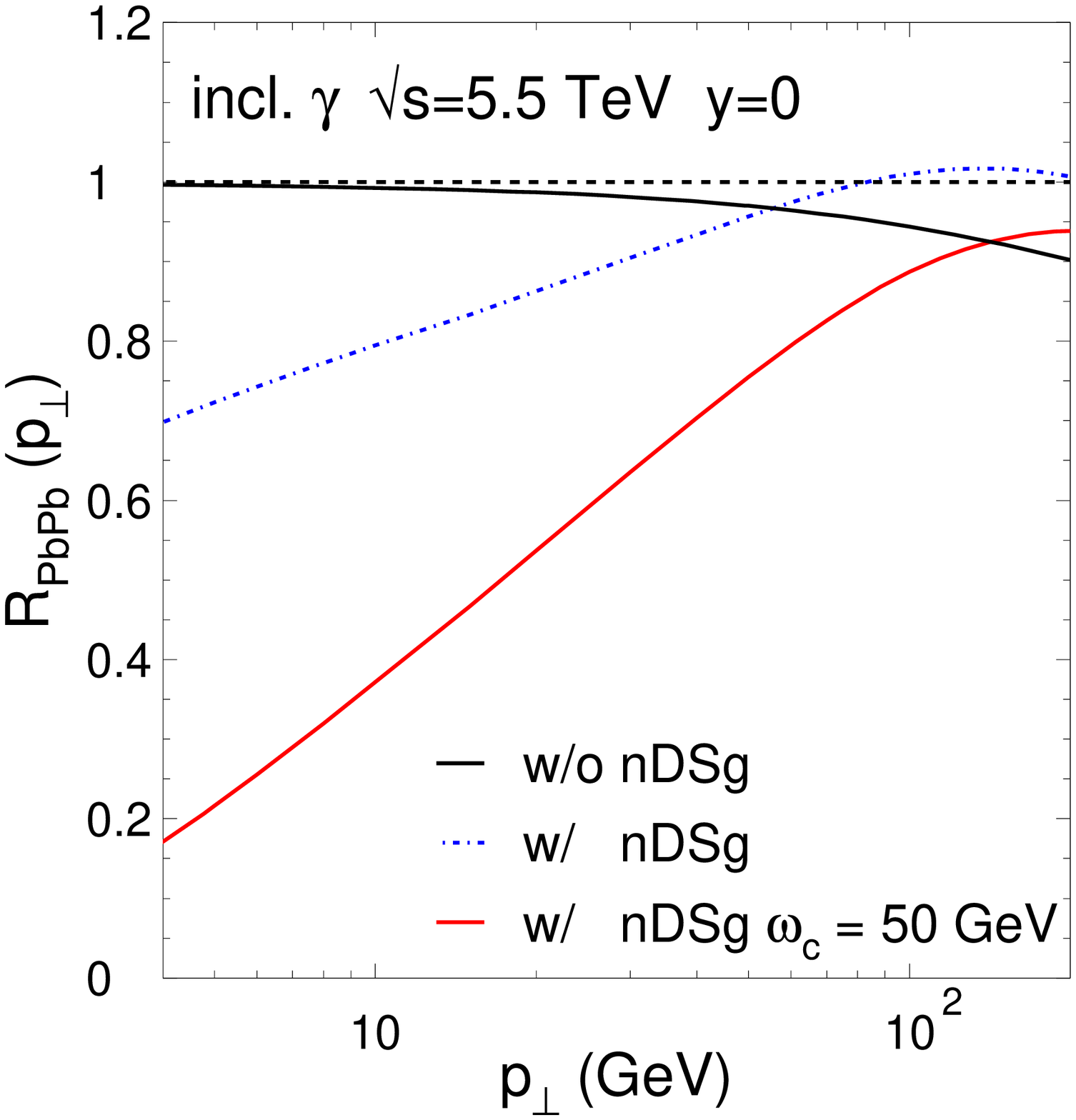}
    \end{center}
  \vspace{-0.5cm}
    \caption{$\rpbpb$ of $y=0$ inclusive photons in Pb--Pb collisions at $\sqrtsnn=5.5$~TeV.}
    \label{fig:incy00lhc}
  \end{minipage}
\end{figure}

Prompt photon production in hadronic collisions has been extensively studied, both experimentally and theoretically, over the past 25 years (see~\cite{Aurenche:2006vj} and references therein). As indicated in Ref.~\cite{Aurenche:2006vj}, it is remarkable that almost all existing data from fixed-target to collider energies can be very well understood within perturbative QCD at NLO. In these proceedings, I briefly discuss how prompt photons in nuclear collisions ($p$--A and A--A) may allow for a better understanding of interesting aspects discussed in heavy-ion collisions, namely the physics of nuclear parton distribution functions and medium-modified fragmentation functions. Parton distribution functions in nuclei are so far poorly constrained, especially in contrast with the high degree of accuracy currently reached in the proton channel, over a wide $x$ and $Q^2$ domain. In particular, only high-$x$ ($x\gtrsim 10^{-2}$) and low $Q^2$ ($Q^2\lesssim 100$~GeV$^2$) have been probed in fixed-target experiments. In order to predict hard processes in nuclear collisions at the LHC, a more accurate knowledge on a wider kinematic range is necessary. As stressed in~\cite{Arleo:2007pc}, the nuclear production ratio of isolated photons in $p$-A collisions,
\begin{equation*}
\label{eq:ratio}
  \rpa(\xt, y) = \frac{1}{A} \ \
  \frac{\dd^3\sigma}{\dd{y}\ \dd^2\ptrans}(p+\A\to\gamma+\X)
  \Big/ \frac{\dd^3\sigma}{\dd{y}\ \dd^2\ptrans}(p+p\to\gamma+\X)
\end{equation*} 
can be related to a good accuracy (say, less than 5\%) to the parton density ratios
\begin{equation*}
  \label{eq:rpa_approx}
\hspace{-2.cm}\rapprox(\xt,y=0) \simeq 0.5\ \rafd(\xt)+ 0.5\ \rag(\xt)\ ;\ \rapprox(\xt,y=3) \simeq \rag(\xt e^{-y}),
\end{equation*}
with $\xt=2\ptrans/\sqrtsnn$. To illustrate this, the ratio $\rpa$ is computed for isolated photons produced at mid-rapidity in $p$--Pb collisions at $\sqrtsnn=8.8$~TeV in Fig.~\ref{fig:isoy0lhc} (solid line), assuming the de Florian and Sassot (nDSg) nuclear parton distributions~\cite{deFlorian:2003qf}. The above analytic approximation $\rapprox_{y=0}$ (dotted line) demonstrates how well this observable is connected to the nuclear modifications of the gluon density and structure function; see also the agreement $(\rpa-\rapprox_{y=0})/\rpa$ as a dash-dotted line in Fig.~\ref{fig:isoy0lhc}. In nucleus-nucleus scattering, the energy loss of hard quarks and gluons in the dense medium presumably produced at LHC may lead to the suppression of prompt photons coming from the collinear fragmentation process. In Fig.~\ref{fig:incy00lhc}, the expected photon quenching in Pb--Pb collisions at $\sqrtsnn=5.5$~TeV is plotted. A significant suppression due to energy loss (taking $\omega_c=50$~GeV, see~\cite{Arleo:2007bg} for details) is observed, unlike what is expected when only nuclear effects in the parton densities are assumed in the calculation (dash-dotted line).


Finally performing momentum correlations between a prompt photon and a leading hadron in $p$--$p$ and A--A collisions, yet experimentally challenging, appears to be an interesting probe of vacuum and medium-modified fragmentation function, as discussed in detail in Refs.~\cite{Arleo:2004xj,Arleo:2006xb}. We refer in particular the interested reader to Fig.~10 of~\cite{Arleo:2004xj} for the predictions of $\gamma$--$\pi^0$ momentum-imbalance distributions at the LHC.

\section*{Acknowledgments}
Part of this work has been done in collaboration with T.~Gousset~\cite{Arleo:2007pc} and P.~Aurenche, Z.~Belghobsi, and J.-P.~Guillet~\cite{Arleo:2004xj}.

\section*{References}

\providecommand{\href}[2]{#2}\begingroup\raggedright\endgroup

\end{document}